# Periodic orbit analysis at the onset of the unstable dimension variability and at the blowout bifurcation

R. F. Pereira [1,2], S. E. de S. Pinto [1], R. L. Viana [2] *, S. R. Lopes [2], and C. Grebogi [3]

1. Departamento de Física, Universidade Estadual de Ponta Grossa,

84032-900, Ponta Grossa, Paraná, Brazil

2. Departamento de Física, Universidade Federal do Paraná,

81531-990, Curitiba, Paraná, Brazil

3. Institute for Complex Systems, King's College,

University of Aberdeen, Aberdeen AB24 3UE, United Kingdom

(Dated: April 12, 2007)

## Abstract

Many chaotic dynamical systems of physical interest present a strong form of non-hyperbolicity called unstable dimension variability (UDV), for which the chaotic invariant set contains periodic orbits possessing different number of unstable eigendirections. The onset of UDV is usually related to the loss of transversal stability of an unstable fixed point embedded in the chaotic set. In this paper, we present a new mechanism for the onset of UDV, whereby the period of the unstable orbits losing transversal stability tends to infinity as we approach the onset of UDV. This mechanism is unveiled by means of a periodic orbit analysis of the invariant chaotic attractor for two model dynamical systems with phase spaces of low dimensionality, and seems to depend heavily on the chaotic dynamics in the invariant set. We also described, for these systems, the blowout bifurcation (for which the chaotic set as a whole loses transversal stability) and its relation with the situation where the effects of UDV are the most intense. For the latter point, we found that chaotic trajectories off but very close to the invariant set exhibit the same scaling characteristic of the so-called on-off intermittency.

---

* Corresponding author. e-mail: viana@fisica.ufpr.br

## I. INTRODUCTION

A chaotic trajectory of a dynamical system is structured upon an infinite set of unstable periodic orbits. A systematic investigation of the latter would provide an understanding of the ergodic properties of the chaotic attractor itself, like its Lyapunov exponents and multifractal spectrum. The relation between chaotic trajectories and unstable periodic orbits embedded therein has been explored in some detail for hyperbolic dynamical systems and non-hyperbolic ones when there are homoclinic tangencies. There is, however, a stronger form of non-hyperbolicity called unstable dimension variability (UDV), which is characterized by the coexistence, in the chaotic invariant set, of periodic orbits with a different number of unstable directions. We harnessed the connection between unstable orbits and chaotic trajectories to investigate the parametric evolution of a chaotic set presenting unstable dimension variability. The onset of UDV is usually related to the loss of transversal stability of an unstable low-period saddle embedded in the chaotic set, through a local bifurcation. We have found, however, a new and qualitatively different scenario underlying the onset of UDV: as we approach this point, the period of the orbits embedded in the chaotic set and losing transversal stability tends to infinity. This fact seems to depend in a sensitive way on the chaotic dynamics in the inveriant set. Another point of interest is the blowout bifurcation point, where the chaotic set as a whole loses transversal stability, the effects of UDV being the most severe in terms of the shadowing properties of computer-generated trajectories. In general, once we have UDV the shadowing distances between fiducial chaotic orbits and noisy ones diverge very fast (i.e. the shadowing times may become small). We made a periodic orbit analysis of the chaotic attractors for two low-dimensional discrete dynamical systems as paradigmatic models, where this new mechanism for the onset of UDV, as well as the blowout bifurcation point, are described. We expect that UDV be rather common in high-dimensional dynamical systems of physical and biological interest, like lattices of coupled oscillators or maps, hence it is important to characterize observable manifestations of the existence of UDV in a given system. Three of them are treated in this paper, in order to provide numerical fingerprints of the scenarios for UDV we have described: (i) the statistics of finite-time Lyapunov exponents (in the transversal direction); (ii) the intermittency exhibited by chaotic orbits off but very close the chaotic set; and (iii) the distribution of log-shadowing distances to the chaotic set. In the first place, at the onset of UDV, the fraction of positive finite-time transversal exponents becomes nonzero, while it takes on the value $1/2$ for the blowout bifurcation point. Secondly, the intermittent switching between a quiescent and a bursting state is influenced by the transversely unstable orbits embedded in the chaotic set, and it has the universal features of the on-off intermittency as we approach the blowout bifurcation point. Finally, we characterized statistically UDV in terms of the exponential distribution of log-(shadowing) distances to the invariant set, and the corresponding power-law scaling exhibited by the corresponding shadowing times.

Unstable periodic orbits (UPOs) are the soul of dynamics, as has been already noted by Poincaré in his work on the restricted three-body problem [1]. More specifically, UPOs form the skeleton upon which the chaotic dynamics is built. In fact, there are infinitely many UPOs embedded in a chaotic attractor. From the ergodic point of view, this description amounts to relate the natural measure generated by a typical chaotic trajectory to the infinite number of atypical measures generated by each of the UPOs embedded in the chaotic attractor [2]. This connection has been rigorously demonstrated for hyperbolic dynamical systems, for which the splitting into stable and unstable invariant subspaces is consistent under the evolution of the dynamics [3]. But it has also been numerically verified for two-dimensional non-hyperbolic systems possessing homoclinic tangencies, which violate the invariant space splitting at the tangency points [4].

There is another class of non-hyperbolic dynamical systems, whose invariant sets also fail to exhibit a consistent invariant subspace splitting. As the trajectory evolves on the chaotic attractor, the dimension of the unstable subspace (and hence the stable) varies densely [5]. In this situation, the breakdown of hyperbolicity is even more severe than when homoclinic tangencies are present, since in the latter case we may found shadowing trajectories for a time often large enough to give meaning to computer-generated single chaotic trajectories [6]. When the unstable dimension varies, however, shadowing trajectories may exist for a time rather short so that in practice one cannot be sure of the validity of the chaotic trajectories numerically generated [7]. For these systems, whose invariant chaotic set is said to exhibit *unstable dimension variability* (UDV), the general relation between natural measure and UPOs is still an open problem.

Two key problems arise in the study of UDV in dynamical systems: the first is its onset, normally related to a bifurcation where some low-period UPO (and its infinitely numerous pre-images) change its unstable dimension [8]. The second problem addresses the point where UDV is the most intense, usually referred to as a blowout bifurcation, since it marks the point at which a chaotic attractor lying in an invariant subspace loses transversal stability as a whole [9–12]. However, not only systems possessing such invariant subspace can exhibit UDV. One counter-example is the kicked double rotor map [13], which was found to exhibit UDV for a limited parameter range, in spite of not having such an invariant manifold embedded in its four-dimensional phase space [14]. One of the numerical fingerprints of UDV is the fluctuations about zero observed for the finite-time Lyapunov exponent (FTLE) closest to zero [15].

In this paper we study the loss of transversal stability of the chaotic invariant set and its relation with the UPOs embedded in it, interpreting our results in terms of the hyperbolicity breakdown caused by UDV. Two archetypical dynamical systems are used to investigate the dynamics: a two-dimensional and a three-dimensional map, possessing invariant subspaces which are chaotic. The blowout bifurcation is interpreted in terms of the parameter value for which the effect of UDV is the most intense [16]. On the other hand, the onset of UDV is described by a mechanism hitherto not considered before: as we approach this point the period of the orbits losing transversal stability tends to infinity. In previous studies [17, 18] the onset of UDV has been described through the loss of transversal stability of a low-period orbit embedded in the chaotic set. We found that the particular scenario for the onset of UDV depends on the chaotic dynamics in the invariant set.

The contribution of the UPOs embedded in the chaotic attractor is numerically analyzed by comparing the fraction of positive transversal finite-time Lyapunov exponents (FTLE) for typical chaotic orbits with the fraction of transversely unstable periodic orbits embedded in the chaotic set [16]. For both system considered in this paper, we find that the onset of UDV and the blowout bifurcation correspond to those points for which this fraction becomes nonzero and it is equal to 1/2, respectively [19].

We also consider the dynamics of chaotic trajectories off but very near the invariant chaotic set. It turns out that they are strongly influenced by the transversal stability of the UPOs embedded in the chaotic set. One manifestation of this influence is a UDV-induced intermittency: the trajectories remain close to the invariant subspace until they approach a transversely unstable orbit and burst away for a while before returning to the vicinity of the invariant subspace [20]. We propose an interpretation of this phenomenon from the point of view of shadowing of chaotic trajectories in the invariant set due to UDV: the shadowing times are the lengths of the laminar intervals for which trajectories remain close to the invariant subspace [21]. The shadowing log-distances and times are found to have exponencial and power-law probability distributions, respectively, in accordance with a biased random-walk model with reflecting barrier [7].

The remaining of this paper is organized as follows: Section II introduces the systems we use to perform the numerical experiments, whereas Section III deals with the computation of the corresponding finite-time Lyapunov exponents and how they can be used to describe UDV. Section IV presents a periodic orbit description of the chaotic attractors focusing on the measures generated by UPOs with different unstable dimensions. Section V discusses the UDV-induced intermittency observed for chaotic trajectories off the invariant set and the statistical properties of log-shadowing distances and shadowing times for such trajectories. Our conclusions are left for the last Section.

## II. DYNAMICAL SYSTEMS AND THEIR PROPERTIES

There are extensive numerical evidence that non-hyperbolic systems with UDV appear typically in prototype models of spatially extended dynamical systems, like coupled maps or oscillator lattices [22]. However, due to the large dimensionality of their phase spaces, a periodic-orbit treatment may become a quite time-consuming task. Hence, we choose to analyze simpler dynamical systems but which retain most of the essential dynamical features expected from those complex lattices.

A typical $N$-dimensional system with such characteristics is [4, 11, 23]

$$\mathbf{x}_{n+1} = \mathbf{f}(\mathbf{x}_n), \tag{1}$$

$$\mathbf{z}_{n+1} = \mathbf{F}(\mathbf{x}_n, \alpha)\mathbf{G}(\mathbf{z}_n), \tag{2}$$

where $\mathbf{x} \in \mathbb{R}^{N_\parallel}$ ($N_\parallel \geq 1$), and $\mathbf{z} \in \mathbb{R}^{N_\perp}$ ($N_\perp \geq 1$) such that $N = N_\parallel + N_\perp$, and $\alpha > 0$ is a bifurcation parameter. We assume the existence of a symmetry in the system so as to warrant the condition $\mathbf{G}(-\mathbf{z}) = -\mathbf{G}(\mathbf{z})$ and, as a consequence, that $\mathbf{G}(0) = 0$, which we interpret as defining a $N_\parallel$-dimensional invariant subspace $\mathsf{S}$ given by $\mathbf{z} = 0$. In terms of a coupled lattice map, for example, $\mathsf{S}$ can stand for the synchronization manifold wherein the time evolution is the same for each map. Our goal of describing the parametric evolution of UDV demands that the dynamics in $\mathsf{S}$ be governed by a map $\mathbf{f}(\mathbf{x}_n)$ with a chaotic attractor. The $N_\perp$-dimensional vector $\mathbf{z}_n$ represents the directions transversal to the invariant manifold, which turn to be important to characterize the stability of the completely synchronized state.

In this paper we analyze two different possibilities for the chaotic dynamics in the invariant subspace [4, 23]:

- Model B, where $N_\parallel = 1$, for which

$$x_{n+1} = 2x_n \;(\text{mod}\; 1), \tag{3}$$

or a Bernoulli shift with a dense chaotic orbit in $[0, 1)$ with Lyapunov exponent $\lambda_x = \ln 2$. In order to generate the numerical trajectories the map was slightly modified to $x_{n+1} = 2x_n + \beta\delta_x \;\text{mod}\; 1$, where $\beta$ is 0 or 1 with equal probabilities and $\delta_x = 2^{-48}$. Due to the transitive property of the strong chaos in $[0, 1)$, the trajectories so generated, although noisy, are shadowed by true chaotic trajectories of the Bernoulli shift map for an arbitrarily long interval [24].

- Model C, where $N_\parallel = 2$, for which

$$x_{n+1} = 2x_n + y_n \;(\text{mod}\; 1), \tag{4}$$
$$y_{n+1} = x_n + y_n \;(\text{mod}\; 1), \tag{5}$$

is the Arnol'd cat map with a hyperbolic chaotic set in the unit square $[0, 1) \times [0, 1)$ [25].

We will consider a one-dimensional transversal dynamics ($N_\perp = 1$) given, for both models, by the following realization of Eq. (2)

$$z_{n+1} = \alpha x_n g(z_n), \tag{6}$$

where we impose that $g(0) = 0$ and $g'(0) = const$. Since $x_n$ has a chaotic behavior, the function $\mathbf{F}(x_n; \alpha) = \alpha x_n$ stands for a driving from the $\mathbf{x}$-dynamics in the invariant subspace to the symmetric **z**-subsystem. We have used the function $g(z) = (1/2\pi)\sin(2\pi z)$, but other maps can be used as well, like the logistic map $g(z) = z(1-z)$.

The Jacobian matrices for models B and C are given, respectively, by

$$\mathbf{DB}(\mathbf{x}_n) = \begin{pmatrix} 2 & 0 \\ (\alpha/2\pi)\sin(2\pi z_n) & \alpha x_n \cos(2\pi z_n) \end{pmatrix}, \tag{7}$$

and

$$\mathbf{DC}(\mathbf{x}_n) = \begin{pmatrix} 2 & 1 & 0 \\ 1 & 1 & 0 \\ (\alpha/2\pi)\sin(2\pi z_n) & 0 & \alpha x_n \cos(2\pi z_n) \end{pmatrix}. \tag{8}$$

The Bernoulli shift map (describing the dynamics in the invariant manifold of map B) has $N_p = 2^p - 1$ unstable periodic points of period $p$ given by

$$x_p(j) = \frac{j}{2^p - 1}, \qquad (j = 0, 1, \ldots 2^p - 2). \tag{9}$$

Each period-$p$ orbit is represented by $x_p(k)$, where $k$ is such that $x_p(k) = \inf\{B^i(x_p(k))\}_{i=0}^{p-1}$. Using this notation the fixed point is $x_1(0) = 0$. When $p = 5$, for example, there are out of $2^5 - 1 = 31$ period-5 points, for example $x_5(1) = \{1/31, 2/31, 4/31, 8/31, 16/31\}$, $x_5(11) = \{13/31, 26/31, 21/31, 11/31, 22/31\}$, and so on.

Similarly, for the cat map, which describes the dynamics in the invariant manifold of C, the counting of the periodic orbits involves $F_p$, the $p$th term of the Fibonacci sequence, given by $F_0 = F_1 = 1$ and $F_p = F_{p-1} + F_{p-2}$ (for $p \geq 2$). The number of period-$p$ (unstable) points is [26]

$$N_p = (F_p + F_{p-2})^2 - 2[1 + (-1)^p], \tag{10}$$

their values being solutions of

$$x_p(j) = A_p x_p(j) + B_p y_p(j) \pmod{1}, \tag{11}$$
$$y_p(j) = C_p x_p(j) + D_p y_p(j) \pmod{1}, \tag{12}$$

where $A_p = F_{2p}$, $B_p = C_p = F_{2p-1}$, $D_p = F_{2p-2}$. These solutions are non-negative rational numbers, for which $x_p(j) = a/c$ and $y_p(j) = b/c$, for some $a, b = 0, 1, 2, \ldots c - 1$ and we have defined [26]

$$c = \begin{cases} F_p + F_{p-2}, & \text{if } p > 1 \text{ is odd,} \\ 5F_{p-1}, & \text{if } p > 1 \text{ is even.} \end{cases} \tag{13}$$

The reason for choosing dynamical systems presenting an invariant subspace in which there is a hyperbolic chaotic set will be clarified in the next sections, where we aim to relate the natural measure of this set with the expanding eigenvalues of the unstable periodic orbits embedded in the chaotic set.

## III. FINITE TIME LYAPUNOV EXPONENTS

The dynamical systems introduced in the last section have some common points of interest that are relevant for our work. Both are skew-symmetric and have invariant subspaces in which the dynamics is strongly chaotic; they are simple enough to allow for a counting of unstable periodic orbits, and they both exhibit the transition to unstable dimension variability through a bifurcation of some periodic orbit. Such a bifurcation changes the unstable dimension of the periodic orbit and its pre-images. As a result of this bifurcation, in the neighborhood of the invariant manifold there appear tongue-like structures anchored at the invariant subspace. The open dense set of tongues are associated with the particular periodic orbit that underwent the bifurcation. The closed set is the stable manifold of the remaining periodic orbits and of the chaotic orbits.

A way to quantify the relative abundance of periodic orbits with a different number of unstable directions is to compute the finite, or time-$n$ Lyapunov exponents of typical chaotic orbits. They are defined, for a $D$-dimensional map $\mathbf{F}(\mathbf{x})$ [27],

$$\lambda_i(\mathbf{x}_0, n) = \frac{1}{n} \ln ||\mathbf{DF}^n(\mathbf{x}_0).\mathbf{e}_i|| \qquad (i = 1, 2, \ldots D), \tag{14}$$

where $\mathbf{e}_i$ is the $i$th singular vector corresponding to the singular value $\xi_i^n$ of the Jacobian matrix $\mathbf{DF}^n$ of the $n$th iterated map. The usual Lyapunov exponents are obtained as the infinite-time limit of the above expression. While, in general, the time-$n$ exponent depends on the initial condition, its infinite-time limit, does not, except for a set of points with Lebesgue measure zero:

$$\lambda_i = \lim_{n \to \infty} \lambda_i(\mathbf{x}_0, n). \tag{15}$$

The eigenvalues of the product of the Jacobian matrices, evaluated at a given point, $\mathbf{DB}^n(\mathbf{x}_0)$, are given by

$$\xi_x^n = 2^n, \qquad \xi_z^n = \alpha^n \prod_{i=0}^{n-1} x_i \cos(2\pi z_i), \tag{16}$$

and, for $\mathbf{DC}^n(\mathbf{x}_0)$, are

$$\begin{aligned}
\xi_x^n &= \frac{1}{2}\left(F_{2n} + F_{2n-2} + F_{2n-1}\sqrt{5}\right), \\
\xi_y^n &= \frac{1}{2}\left(F_{2n} + F_{2n-2} - F_{2n-1}\sqrt{5}\right), \\
\xi_z^n &= \alpha^n \prod_{i=0}^{n-1} x_i \cos(2\pi z_i),
\end{aligned} \qquad (17)$$

The characterization of non-hyperbolicity *via* unstable dimension variability is based on the finite-time exponent closest to zero. For our low-dimensional maps, this turns to be the exponent computed along the transversal direction to the invariant manifold, which we henceforth will refer to as $\lambda_\perp(n)$. The time-$n$ transversal exponent for both models, according to Eqs. (16) and (17), is given by

$$\lambda_\perp(n) = \frac{1}{n}\sum_{i=0}^{n-1} \ln(\alpha x_i), \qquad (18)$$

such that the transversal exponent is the infinite-time limit of the above expression:

$$\lambda_\perp = \lim_{n\to\infty} \lambda_\perp(n) = \int_0^1 \ln(\alpha x)\rho(x)dx = \ln\alpha - 1, \qquad (19)$$

where we have used the Poincaré ergodic theorem and $\rho(x) = 1$ is the invariant density of iterations in the unit interval for model B or its two-dimensional equivalent for the unit square in model C. It follows from Eq. (19) that the blowout bifurcation, i.e., the point where the invariant manifold becomes transversely unstable as a whole, occurs at $\alpha^* = e$ for both models.

A fingerprint of unstable dimension variability (UDV) is the existence of positive and negative fluctuations for $\lambda_\perp(n)$. This means that the transversal exponents along a chaotic trajectory typically present finite-time segments for which the phase-space volumes are locally either expanding or shrinking on average. We take a long chaotic trajectory and partition it into a large number of segments with length $n$. After computing the time-$n$ exponents for each trajectory segment, we are able to obtain a frequency distribution, related to a probability density $P(\lambda_\perp(n))$. For $n$ large enough this distribution has a Gaussian-like shape, whose width decreases as $\sim n^{-1/2}$ such that, in the limit $n \to \infty$ it reduces to a delta-function centered at the infinite-limit $\lambda_\perp$ [19]. It should be mentioned, however, that UDV is not necessarily related with the non-Gaussian nature of the probability distributions but with the fact that they have tails of positive finite-time exponents.

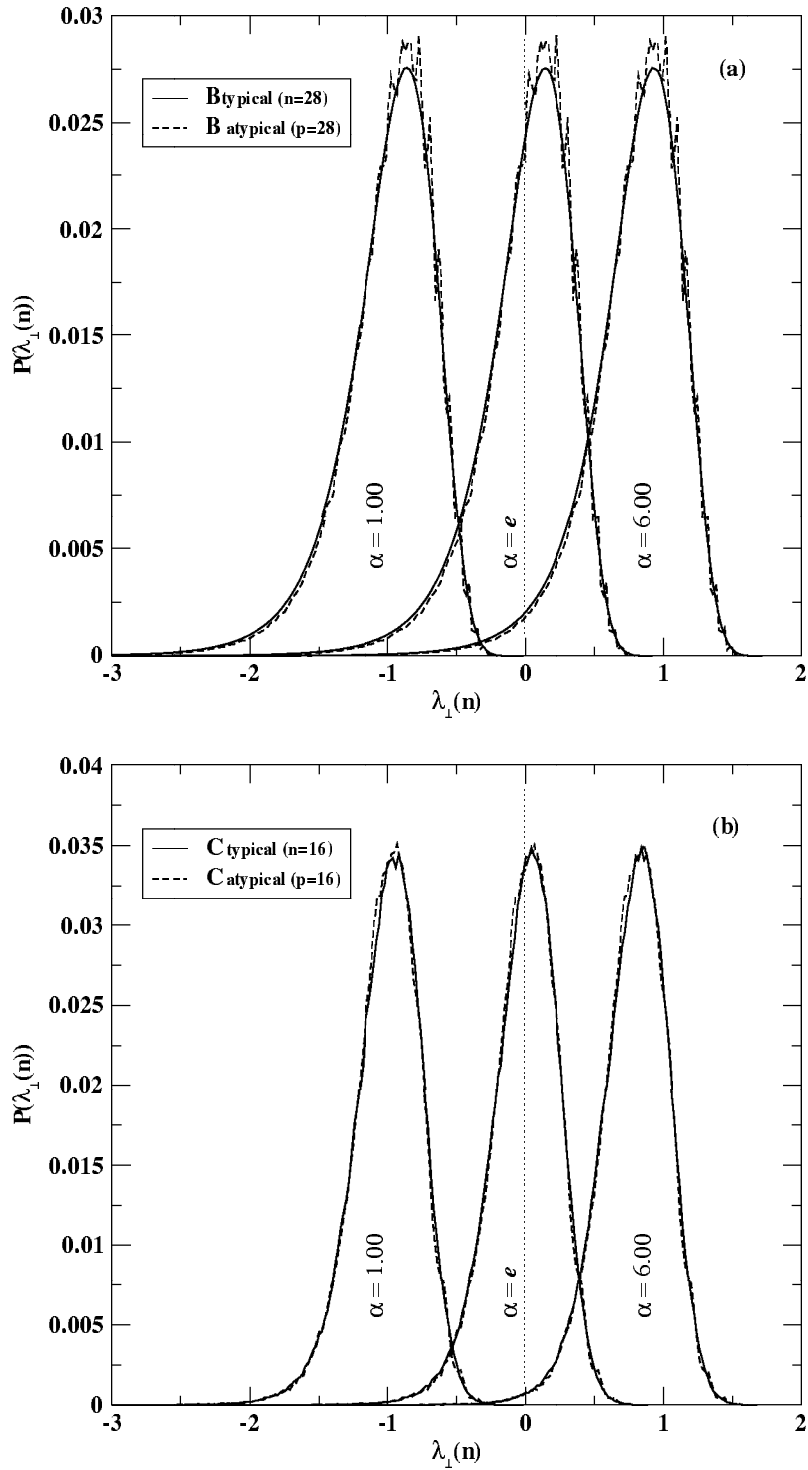

FIG. 1: Probability distribution of time-$n$ transversal Lyapunov exponents for maps (a) B, $n = 28$; and (b) C, $n = 16$, and different values of the bifurcation parameter $\alpha$. Solid lines: distributions evaluated using typical chaotic trajectories with $n \times N_n$ points. Dashed lines: distributions computed using all unstable period-$n$ orbits.

| TABLE I: Model B | | |
|---|---|---|
| $\alpha$ | $\lambda_{max}$ | $\lambda_\perp$ |
| 1.00 | - 0.86 | -1.00 |
| 2.72 | 0.14 | 0.00 |
| 6.00 | 0.79 | 0.92 |

| TABLE II: Model C | | |
|---|---|---|
| $\alpha$ | $\lambda_{max}$ | $\lambda_\perp$ |
| 1.00 | - 0.99 | -0.96 |
| 2.72 | 0.00 | 0.06 |
| 6.00 | 0.76 | 0.84 |

Numerical approximations to this probability density for time-$n$ transversal exponents are shown as solid lines in Figures 1(a) and (b) for models B ($n = 28$) and C ($n = 16$), respectively. The dashed line represents the analogous distribution for all UPOs of period-$p$ ($p = n$) embedded in the chaotic attractor, which nearly coincides with the distributions obtained from the typical chaotic trajectories. We found both distributions to have a Gaussian-like shape, drifting without noticeable distortion towards positive exponents as the control parameter $\alpha$ increases [28]. Actually the distributions have uneven tails, the positive tail having a cutoff at the point $\lambda_\perp(n) = \ln \alpha$, whereas the negative tail looks like a Gaussian one. If the distributions were Gaussians their maxima $\lambda_{max}$ would coincide with the infinite-time exponent $\lambda_\perp$ given by Eq. (19). This is not the case for the distributions shown in Fig. 1, though, as can be seen in Tables I and II where we present the corresponding values of $\lambda_{max}$ and $\lambda_\perp$ (obtained from Eq. (19), which are clearly different.

There are two critical points to be explored in the evolution of the distribution $P(\lambda_\perp(n))$ for increasing $\alpha$: (i) if $\alpha < \alpha_c \approx 1.0$ there are no positive fluctuations of the finite-time exponent, indicating absence of UDV. The critical value $\alpha_c$ signals the onset of UDV, since positive fluctuations start appearing at this point due to the existence of transversely unstable UPOs embedded in the attractor; (ii) for $\alpha_c < \alpha \lesssim \alpha^* = e$, there are more negative than positive fluctuations, indicating that the distribution peak lies in the negative semi-axis. Accordingly, before the blowout bifurcation the infinite-time exponent $\lambda_\perp$ is likewise negative, but, at the blowout bifurcation $\alpha^* = e$, the distribution peak is slightly displaced from zero, since the distributions themselves have unequal tails due to non-gaussianity. (iii) When $\alpha > \alpha^*$ the positive fluctuations dominate.

Another way to understand these observations is to compute the positive fraction of the transversal finite-time exponent, or

$$\phi(n) = \frac{\int_0^{+\infty} P(\lambda_\perp(n)) d\lambda_\perp(n)}{\int_{-\infty}^{+\infty} P(\lambda_\perp(n)) d\lambda_\perp(n)}. \tag{20}$$

The solid lines depicted in Figures 2(a) and (b) present the computed fraction of time-$n$ transversal exponents for maps B ($n = 28$) and C ($n = 16$), respectively, as a function of the bifurcation parameter $\alpha$. Since this fraction exhibits a monotonically increasing trend with $\alpha$, we can use the condition $\phi > 0$ for characterizing the onset of UDV in those systems, giving an estimated value of $\alpha_c \approx 1.0$ for this critical point. However, the insets in Fig. 2 reveal small fluctuations near this critical point (due to the finite bin width) making difficult a precise evaluation of $\alpha_c$ based solely on this criterion. This is not surprising, however, since the characterization of UDV through finite-time Lyapunov exponents is difficult, inasmuch, we must deal with exponents fluctuating about zero, and at this parameter value the density of the attractor is small so that very long trajectories are needed to obtain reliable results. In order to overcome such problems, a numerical method based on the scaling of the probability distribution of points near the fixed points has been recently proposed [29].

At the blowout bifurcation, there follows from Fig. 2 that nearly half of the finite-time transversal exponents are positive: this fraction actually takes on values around 0.57 and 0.54 for models B and C, respectively. The difference with the theoretical value 0.5 is due to the use of a finite (and small) time interval $n$ in the computation of the probability distribution $\lambda_\perp(n)$. We remark that we have used such small values of $n$ since we set $n$ equal to the period $p$ of the UPO's used as examples of atypical orbits. We are particularly interested to explore the situation that lies between the onset of UDV, for which the first UPO loses transversal stability, and the blowout bifurcation, where the manifold as a whole becomes transversely unstable. However, this does not necessarily mean that for the blowout bifurcation all UPOs have lost transversal stability. The answer lies in the relative contribution of the two sets of UPOs with different unstable dimensions.

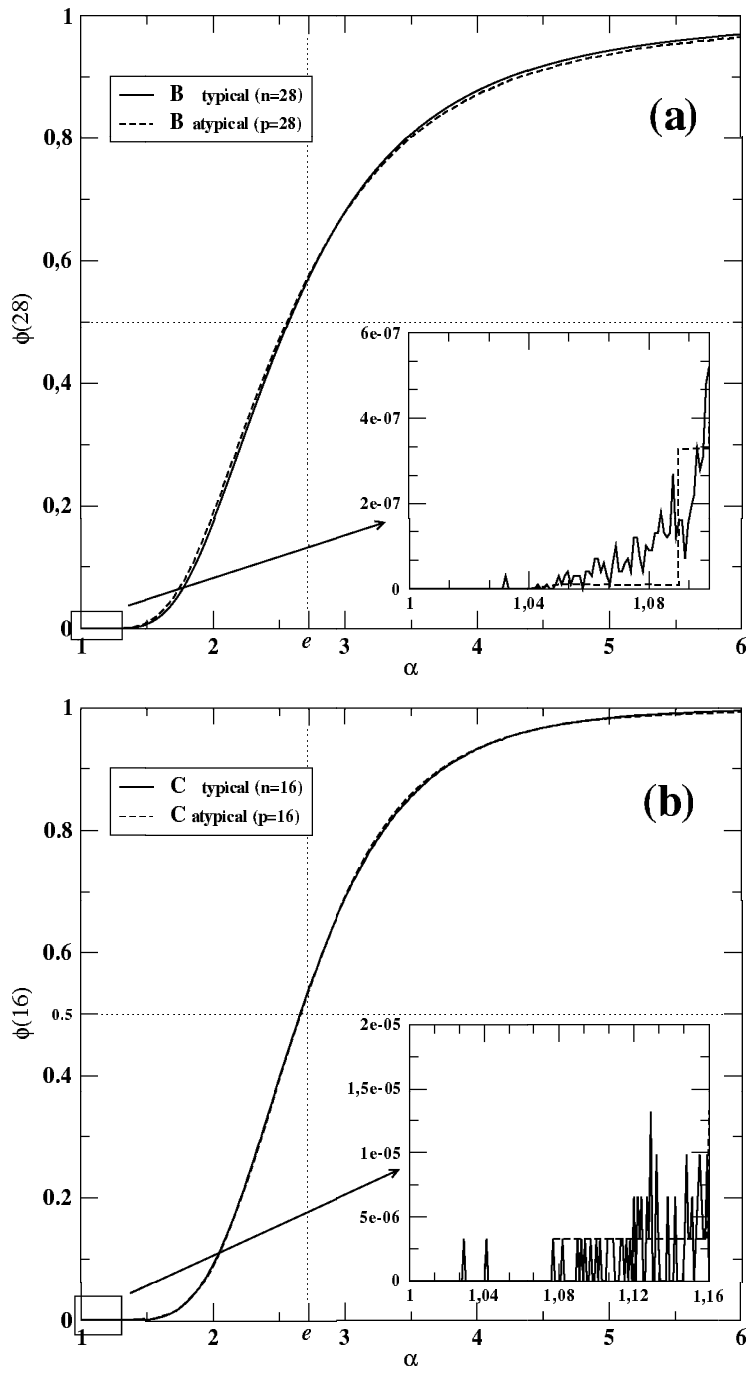

FIG. 2: Fraction of positive time-$n$ transversal Lyapunov exponents for maps (a) B ($n = 28$) and (b) C ($n = 16$) as a function of the bifurcation parameter $\alpha$. Solid lines: typical chaotic orbits; dashed lines: fraction of transversely unstable period-$p$ orbits. The insets in both figures show blowouts of the regions near the onset of UDV for both maps.

## IV. TRANSVERSAL STABILITY OF UNSTABLE PERIODIC ORBITS

After the onset of UDV ($\alpha \gtrsim \alpha_c$) there are two kinds of coexisting UPOs: saddles and repellers. The saddles (repellers) are transversely stable (unstable) and thus have unstable dimension one (two). Just after the onset of UDV, the contribution of repellers to the measure of the chaotic set of the invariant subspace is very weak, since the UPO which lost transversal stability has Lebesgue measure zero. As one approaches the blowout bifurcation at $\alpha^*$ the contribution of repellers becomes increasingly significant until it matches the contribution of the saddles. After the blowout bifurcation point, the contribution of repellers to the measure become dominant.

The relative contributions from saddles and repellers in the context of UDV can be made more precise in the context of periodic-orbit theory [11, 12]. In order to compute the infinite-time transverse Lyapunov exponent $\lambda_\perp$, we use *typical* trajectories on the chaotic invariant set $\Omega$, with respect to its natural measure $\mu(\Omega)$ ($\Omega$ is the unit interval $[0,1]$ and the unit square for models B and C, respectively). Since there are an infinite number of UPOs embedded in $\Omega$, they support the natural measure in the sense that, when computing $\lambda_\perp$, these periodic orbits contribute with different weights. These weights are given by the natural measure of a typical trajectory which visits the neighborhoods of the periodic orbits, and are related to the magnitude of the unstable eigenvalues of those unstable orbits [3].

The influence of the UPO's on the loss of transversal stability of the chaotic invariant set can be appreciated in Fig. 3(a), where we plotted the critical value of the bifurcation parameter $\alpha_{first}(p)$ for which the *first* UPO of period $p$ loses transversal stability, as a function of the period $p$ of the UPO. The value of $\alpha_{first}(p)$ decreases monotonically for increasing period and seems to asymptote to 1 for large $p$, this evolution being essentially the same for models B and C; a result which is in close agreement with the estimate we made from Fig. 2. These observations can be made rigorous by a straightforward stability calculation we present in detail in an Appendix. We conclude that the first orbits to lose transversal stability are those with very large periods. In this sense, the atypical trajectories represented by UPOs do not dominate the effects produced by typical trajectories in the chaotic invariant set.

Figure 3(b) depicts the values of $\alpha$ for which the *last* UPO of period-$p$ loses transversal stability for models B (circles) and C (crosses). The solid lines are least squares fits of the form $\exp(\psi p)$, where $\psi = 0.35$ and $0.26$ for models B and C, respectively. If we consider a fixed value of $\alpha$, say $\alpha = 2$, it would seem that most UPOs (mainly with large period) have already lost transversal stability. However, Fig. 3(b) shows that, at $\alpha = 2$, there are still transversely stable UPOs of all periods. The conclusion we get is that large-period UPOs lose transversal stability earlier (in terms of increasing $\alpha$-values) than low-period ones, but it takes longer to exhaust all possibilities.

This can be viewed as a consequence of the number of period-$p$ orbits to increase with their period $p$ as $e^{h_T p}$, where $h_T > 0$ is the topological entropy of the chaotic set, such that there are far more UPOs with large periods than low-period ones [30]. A standard calculation shows that $h_T = \ln 2 \approx 0.69$ for model B, and

$$h_T = \lim_{n \to \infty} \ln \left( \frac{F_n}{F_{n-1}} \right)^2 \approx 0.97 \qquad (21)$$

for model C. Hence the number of period-$p$ orbits increases more rapidly for model C than for model B, such that, for a fixed $p$, it is more difficult to count these orbits for model C. For this reason, we have limited ourselves to UPOs of periods up to 28 and 16 when considering models B and C, respectively. Moreover, we also expect predictions from periodic orbit theory to agree more with numerical results for model B.

The fact that the onset of UDV is marked by the loss of transversal stability of a UPO with infinite period seems to depend in a sensitive way on the chaotic dynamics in the invariant set, keeping the same transversal dynamics. For example, when we used in model B, a tent map $x_{n+1} = 1 - 2|x_n - (1/2)|$ (or its topological conjugate, $x_{n+1} = 4x_n(1 - x_n)$) instead of the Bernoulli shift map, the first orbit to lose transversal stability is a fixed point.

The natural measure of a typical trajectory in the neighborhood of a periodic orbit is related to the probability of being in its close vicinity, which is smaller the more unstable the periodic orbit is [2]. Hence UPOs with large unstable eigenvalues have a comparatively smaller contribution to the natural measure. Summing over all unstable period-$p$ orbits embedded in the invariant set $\Omega$ gives then its natural measure [3]

$$\mu(\Omega) = \lim_{p \to \infty} \sum_{\mathbf{x}_p(j) \in \Omega} \frac{1}{L_u(\mathbf{x}_p(j))}, \qquad (22)$$

where $\mathbf{x}_p(j)$ is the $j$th fixed point of $\mathbf{F}^p(\mathbf{x})$ belonging to the set $\Omega$, i.e., $\mathbf{x}_p(j)$ is on a period-$r$ orbit, where $r$ equals either $p$ or a prime factor of $p$.

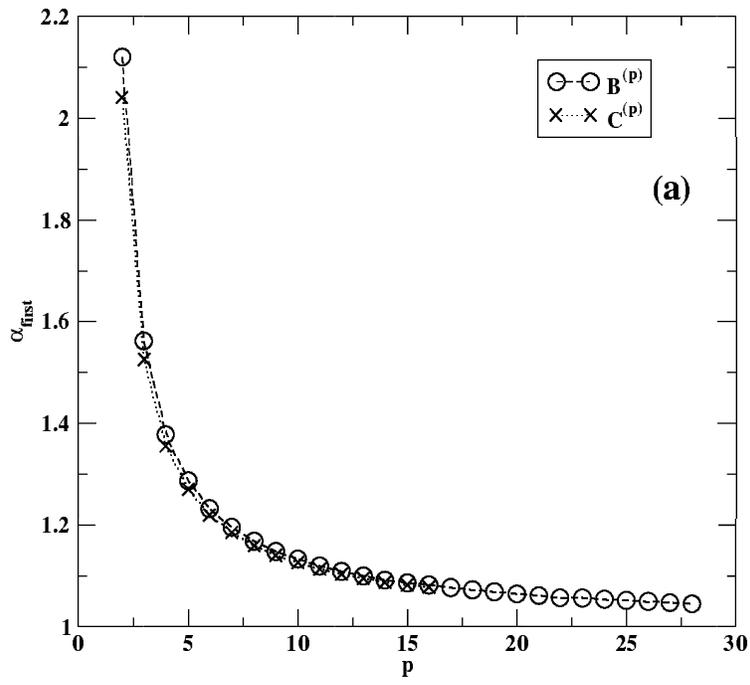

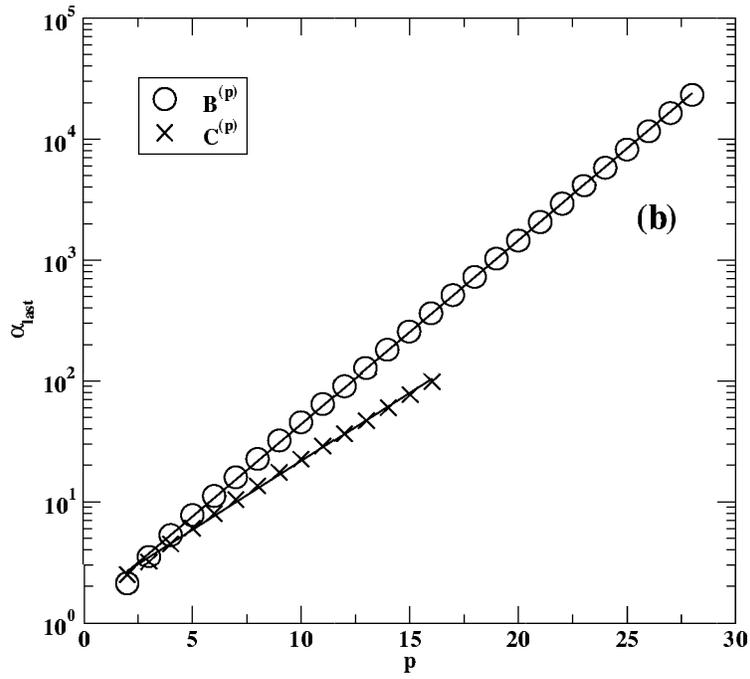

FIG. 3: Critical values $\alpha_p$ of the bifurcation parameter for which the (a) first and (b) last period-$p$ orbits lose transversal stability. Circles and crosses refer to models B and C, respectively. The lines in (b) are least squares fits with slopes 0.35 and 0.26 for models B and C, respectively.

According to Ref. [3], $L_u$ is the absolute value of the product of eigenvalues related to the unstable directions of $\mathbf{DF}^p$, computed for the orbit points $\mathbf{x}_p(j)$:

$$L_u(\mathbf{x}_p(j)) = \left| \prod_{i=1}^{D} \xi_i(\mathbf{x}_p(j)) \right|, \qquad (23)$$

such that $|\xi_i| > 1$ and $D$ is the dimensionality of the map $\mathbf{F}$. For the models treated in this paper, it turns out that $L_u$ does not depend on $\mathbf{x}_p(j)$ since the $x$-direction is always unstable and $\xi_x$ takes on the same values for all orbits $\mathbf{x}_p(j)$, such that $L_u = 2^p$ for model B, and $L_u = (1/2)(F_{2p} + F_{2p-1} + \sqrt{5}F_{2p-2})$ for model C.

Equation (22) was originally derived for hyperbolic chaotic sets, for which there are Markov partitions formed by intersections of stable and unstable manifolds of UPO embedded in the chaotic invariant set $\Omega$ [3]. In subsequent papers [4, 11] non-hyperbolic systems with homoclinic tangencies between stable and unstable manifolds of unstable periodic orbits were considered, and Eq. (22) was used by assuming that $L_u$ would be the absolute value of the greater eigenvalue of the orbit $\mathbf{x}_p(j)$, or, for model C,

$$L_u(\mathbf{x}_p(j)) = \sup\{|\xi_x^p|, |\xi_y^p|, |\xi_z^p|\}. \qquad (24)$$

In many instances, Eqs. (23) and (24) would give the same result for the natural measure of typical chaotic orbits, $\mu(\Omega)$, provided one computes *only the eigenvalues related to unstable directions within the invariant set* $\Omega$. The results of Refs. [4, 11] did not differ from the original definition of $L_u$, Eq. (23), because there was only one unstable direction for orbits in the chaotic set $\Omega$. On the other hand, higher dimensional systems such as the ones we deal with in this paper, there can be unstable directions *not in the chaotic invariant set* $\Omega$, namely those related to transversal directions. Since these chaotic sets are taken to be hyperbolic by construction, Eq. (22) can be applied to them, but there may be differences between the results furnished by Eqs. (23) and (24). However, since an unstable direction (the $x$-direction) along the invariant subspace has usually the greater eigenvalue, the possible discrepancies between (23) and (24) are conceptually significant though small in size as we shall show.

The (atypical) natural measure associated with the $j$th period-$p$ orbit is the normalized ratio [16]

$$\mu_p(j) = \frac{1/L_u(\mathbf{x}_p(j))}{\sum_{\ell=1}^{N_p}[1/L_u(\mathbf{x}_p(\ell))]}, \qquad (25)$$

where $N_p$ is the number of period-$p$ orbits. $N_p^s$ and $N_p^u$ are the numbers of transversely stable and unstable period-$q$ orbits, respectively, such that $N_p^s + N_p^u = N_p$. The weights of the transversely stable and unstable period-$q$ orbits are given, respectively, by

$$\Lambda_p^s = \sum_{N_p^s} \mu_p(j) \lambda_\perp(\mathbf{x}_p(j), p) \qquad (\text{for} \quad \lambda_\perp(\mathbf{x}_p(j), p) < 0), \tag{26}$$

$$\Lambda_p^u = \sum_{N_p^u} \mu_p(j) \lambda_\perp(\mathbf{x}_p(j), p) \qquad (\text{for} \quad \lambda_\perp(\mathbf{x}_p(j), p) > 0), \tag{27}$$

where $\lambda_\perp(\mathbf{x}_p(j), p)$ is the time-$p$ transversal Lyapunov exponent for the $j$th period-$p$ orbit. If $\lambda_\perp(\mathbf{x}_p(j), p)$ is positive (negative) the periodic orbit is transversely unstable (stable).

Within the realm of the periodic-orbit theory, when $\lambda_\perp = 0$ (blowout bifurcation), it follows that the contributions of the transversely stable and unstable period-$q$ orbits are exactly matched. In this sense, we can say that UDV is the most intense at blowout bifurcation, and the validity of computer-generated chaotic trajectories is rather short since there would be no shadowing trajectories of the chaotic invariant set for a significant lifespan [31]. In terms of the finite-time transverse exponents, one could expect that half of the time-$n$ exponents would be positive, which would imply that, at the blowout bifurcation, $\phi(n) = 1/2$. While this connection has been numerically verified for some dynamical systems [19], an inspection of Fig. 2 reveals discrepancies with this assertion, for the blowout bifurcation occurs slightly *after* $\phi$ equals 0.5.

Another way of analyzing the relative contribution of period-$q$ saddles and repellers is to use the so-called contrast measure, defined as

$$C_p = |\mu_p^u - \mu_p^s|, \tag{28}$$

where

$$\mu_p^s = \sum_{i=1}^{N_p^s} \mu_p(i), \quad \mu_p^u = \sum_{i=1}^{N_p^u} \mu_p(i) = 1 - \mu_p^s, \tag{29}$$

and $\mu_p(i)$ is given by Eq. (25). The variation of the contrast measure with $\alpha$ is depicted in Fig. 4 for UPOs with period-$p$. Before the onset of UDV ($\alpha \lesssim \alpha_c$) the only contribution is that from the saddles, hence $C_p = 1$. The vanishing of $C_p$ marks the point where the relative contributions of period-$p$ saddles and repellers exactly match each other. For both models, this point does not coincide with the blowout bifurcation at $\alpha^* = e$, although for model C the difference is smaller than for model B, in agreement with the results obtained for the fraction of positive finite time exponents in Fig. 2. The explanation for this small discrepancy lies in the use we made of finite and small periods for the UPOs to which atypical measures are compared. We expect periodic-orbit predictions to match better with numerical results if $p$ is enlarged.

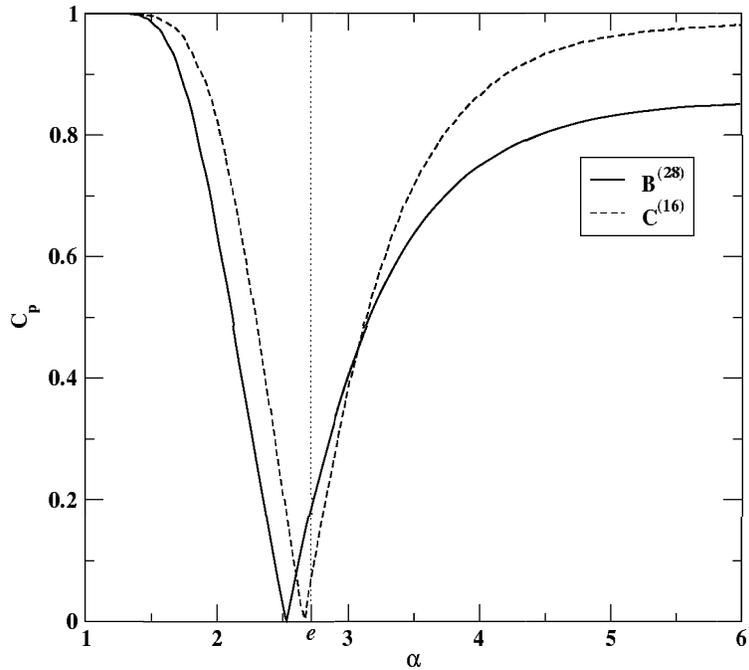

FIG. 4: Contrast measure *versus* bifurcation parameter for unstable orbits with period: (a) $p = 28$ (model B); (b) $p = 16$ (model C).

Although analytical methods to count unstable periodic orbits exist only for low-dimensional maps, such as those studied in this paper, there are powerful and efficient numerical methods devised to do this task even for more complex systems [32–34] The problem with those methods is that the computational time required grows both with the period of the orbit as well as with the dimension of the dynamical system. For a $D$-dimensional map, the method proposed in Ref. [34] requires that, in order to find orbits of period $p + 1$, out of $2^{d_u}$ recursive equations (involving $D \times D$ matrices) be solved for each period-$p$ orbit with unstable dimension $d_u$. But the number of periodic points grows exponentially fast according to the topological entropy. As a representative example of high-dimensional dynamical system, for the kicked double rotor map [13] there are more than $300,000$ period-7 points with unstable dimensions equal to 1 or 2, such that one must deal with approximately a million of matrix equations to find the period-8 orbits. On the other hand, since in the scenario immediately following the onset of UDV we need orbits of arbitrarily large period, the use of such numerical methods might not be feasible to find the value of $\alpha_c$. That was essentially the reason behind our using of simple, low-dimensional maps for which a direct analytical treatment is feasible. In order to establish the existence of such scenario in more complex systems, like higher-dimensional flows or coupled maps, one would need to resort to those time-consuming methods.

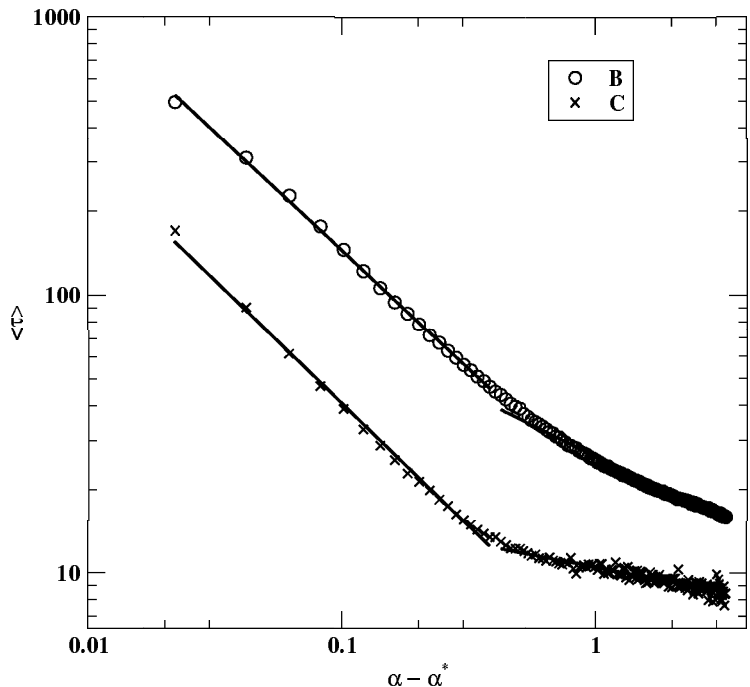

FIG. 5: Average number of iterations for a typical trajectory of model B (circles) and C (crosses) to be trapped in a vicinity of width $10^{-12}$ of the chaotic set *versus* the difference between the bifurcation parameter and its critical value $\alpha^* = e$. The solid lines are least squares fits, cf. Eq. (30), with slopes $-0.852$ and $-0.447$ for model B, and $-0.885$ and $-0.177$ for model C.

## V. UDV-INDUCED INTERMITTENCY

The existence of UDV in the invariant chaotic set $\Omega$ for the models considered in this work, has an observable effect on trajectories which do not belong exactly to the chaotic set but are near enough to feel the coexistence of periodic orbits transversely stable and unstable. These trajectories have typically an intermittent behavior, alternating quiescent and bursting phases. We can call the process *UDV-induced intermittency*, since here chaotic bursting is accompanied by the lack of hyperbolicity [20].

A trajectory which is initialized off but very close to the invariant chaotic set $\Omega$ will be pushed either away or towards $\Omega$ if it is in the vicinity of a repeller or a saddle UPO, respectively. The time it takes for such a trajectory to leave a small neighborhood of $\Omega$, say, of size $10^{-12}$, diverges rapidly as we are close to the blowout bifurcation, as shown by Fig. 5, where we observe two scaling regions for which

$$<\tau> \sim |\alpha - \alpha^*|^{-\varpi} \qquad (30)$$

where the slopes are different for the region near and far the critical point, with a shoulder at $|\alpha - \alpha^*| = O(1)$. As a consequence, if we are interested to investigate the behavior of long trajectories in the vicinity of blowout bifurcation, it would take a prohibitively long time unless we add an extrinsic noise $\gamma R_n$ in the transversal dynamics of both models B and C, where $\gamma$ is a noise level of $10^{-16}$ and $R_n$ is a pseudo-random variable in the interval $[0, 1]$ [7, 21].

This procedure does not alter the transversal dynamics in the scenario of UDV, since we can regard the transversal finite-time Lyapunov exponents as random kicks that push the trajectory away or towards $\Omega$ as the exponents are positive or negative, respectively. Hence we are just adding an extra noisy term to an already noisy system (the transversal dynamics near $\alpha^*$, where the positive and negative finite-time Lyapunov exponents match each other) that does not alter the difference between up and down kicks due to UDV. However, this procedure should not be used too far from the blowout bifurcation, otherwise the distribution of kicks due to the finite-time exponents is biased towards either positive or negative value, and adding a white noise with uniform distribution would lead to artificial results.

Representative examples of intermittent behavior induced by UDV are depicted in Figs. 6(a) and (b) for models B and C, respectively, and a value of $\alpha$ very close to the blowout bifurcation: $\alpha = \alpha^* + \delta$, with $\delta \sim 10^{-3}$. In both cases we can define a laminar phase as the inter-burst interval with duration $\tau$. As in most intermittent scenarios, the duration of these laminar phases differ greatly but they are statistically distributed according to a power-law scaling [cf. Fig. 7, where the scaling was obtained through a least-squares fit] $P(\tau) \sim \tau^{-\omega}$, where $\omega$ was estimated to take on the values 1.492 and 1.480 for maps B and C, respectively. Both values compare well with the theoretical value 3/2 predicted for on-off intermittency

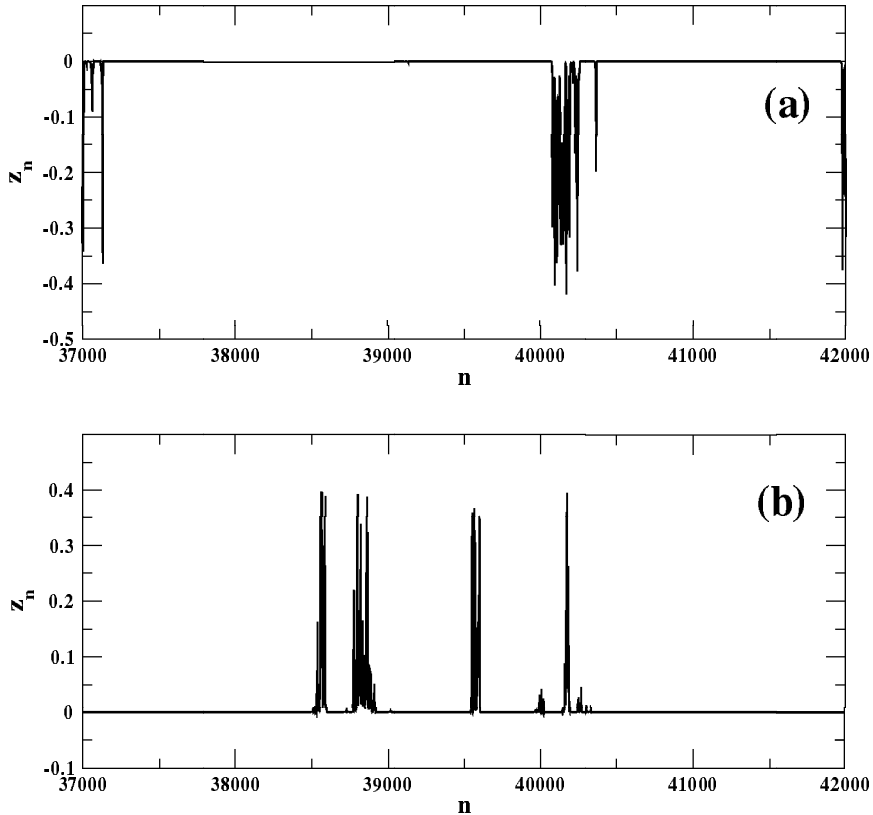

FIG. 6: Evolution of the transversal coordinate of a trajectory off but very close to the chaotic set of models (a) B and (b) C, with $\alpha = 2.72 \gtrsim \alpha^*$.

[35].

The connection between UDV and on-off intermittency has been pointed out in Refs. [36], where it is claimed to be a universal feature. In fact, the observed intermittent transition to complete synchronization of coupled maps satisfies the same power-law scaling with 3/2 exponent [18, 31]. In the latter case, the invariant chaotic set is the synchronization manifold. As the coupling strength varies over a given interval, the coupled map lattice presents UDV. The blowout bifurcation corresponds to the situation whereby the synchronization manifold loses transversal stability. A non-synchronized trajectory very close to this manifold experiences the same kind of kicks towards and away the synchronization manifold. The 3/2-exponent comes from the fact that the on-off intermittency comes from a system parameter being perturbed in the vicinity of a bifurcation point (in our case the blowout bifurcation). Our results provide further numerical evidence that the behavior of chaotic transients near a blowout bifurcation do present features universal to on-off bifurcation.

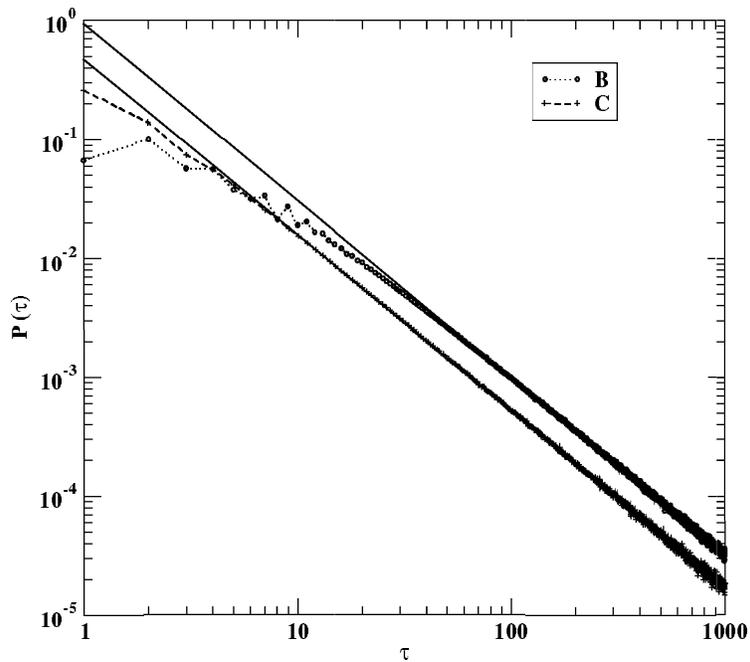

FIG. 7: Average inter-burst interval for models B and C with $\alpha = \alpha^* + \delta$, $\delta = 10^{-3}$. Solid lines represent least-squares fit with slopes $-1.492$ and $-1.480$, respectively.

The UDV-induced intermittency can also be viewed from the point of view of the shadowability of trajectories in the chaotic invariant set $\Omega$. In practice, due to unavoidable factors as imperfect parameter identification and one-step truncation errors involved in the numerical procedures, a computer-generated trajectory thought to belong to $\Omega$ will actually start off but very close to it. However, we know in advance that there exists a "true" chaotic trajectory in $\Omega$, which may or may not shadows the pseudo-trajectory obtained through numerics.

The shadowing distance between the "true" chaotic trajectory at and the pseudo-trajectory initialized nearby is, at each instant, the pointwise distance between them in the phase plane. We usually compute the log-distance $d_n = \log_{10} z_n$ (cf. Fig. 8) for the time-series of this variable for the same case depicted in Fig. 6. The existence of laminar intervals, for which the pseudo-trajectory is close to $\Omega$, is equivalent to having a pseudo-trajectory which continuously shadows the "true" chaotic trajectory belonging to $\Omega$. Bursting is an observable manifestation of the lack of shadowability, while the lengths of the laminar intervals yield estimates for shadowing times. Hence, the properties of chaotic bursting are related to the statistics of shadowing distances and times.

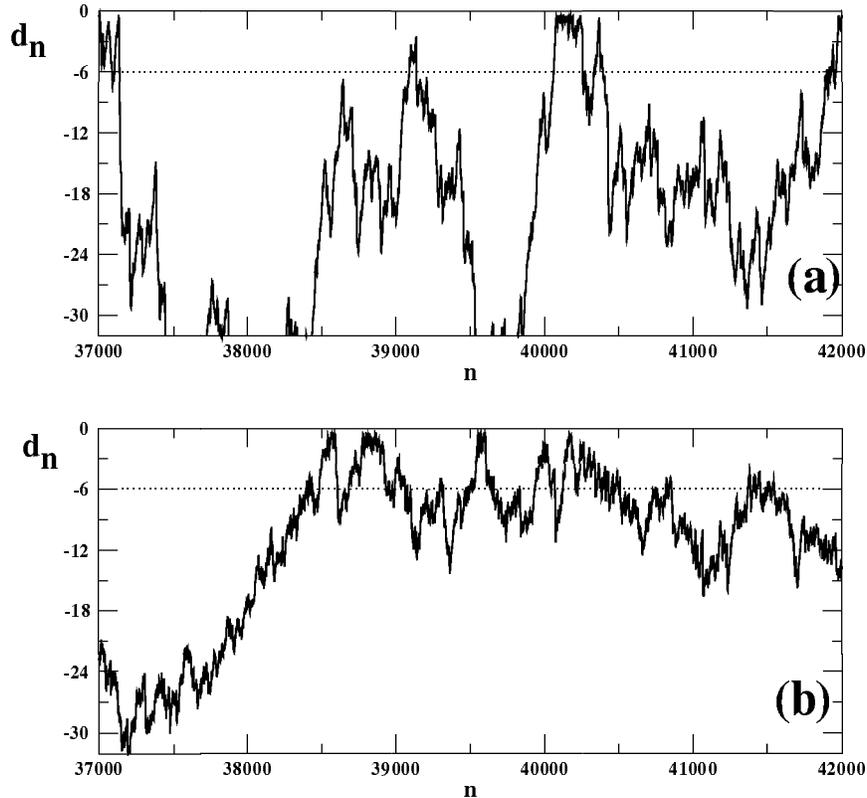

FIG. 8: Evolution of the log-shadowing distances for the same time series depicted in Figs. 6(a) and (b).

Figures 9(a) and (b) present numerically obtained statistical distributions of the shadowing log-distances $d_n$ for models B and C, respectively, and different values of $\alpha$, where we applied external kicks of level $\delta_\perp = 10^{-16}$ as before. When $\alpha$ is far from the blowout bifurcation (solid and dashed lines in Fig. 9) the (normalized) distribution height falls rapidly down to zero for shadowing distances less than $10^{-16}$, as expected due to the noisy applied kicks, and decreases exponentially for higher shadowing distances $\mathcal{P}(d) \sim \exp(-\kappa d)$, where $\kappa > 0$. As $\alpha$ increases towards the blowout bifurcation, the slowdown rate $\kappa$ decreases until it becomes practically zero near the bifurcation (dotted lines in Fig. 9). This can be interpreted from the fact that as the UDV effect is more intense near the blowout bifurcation, we have a progressive dominance of higher shadowing distances. This is in accordance with the greater content of transversely unstable periodic orbits as $\alpha$ is increased further from $\alpha^*$.

The shadowing log-distances experience spikes of various heights, but remain in the immediate vicinity of the invariant subspace $\Omega$, until they burst chaotically and return to $\Omega$.

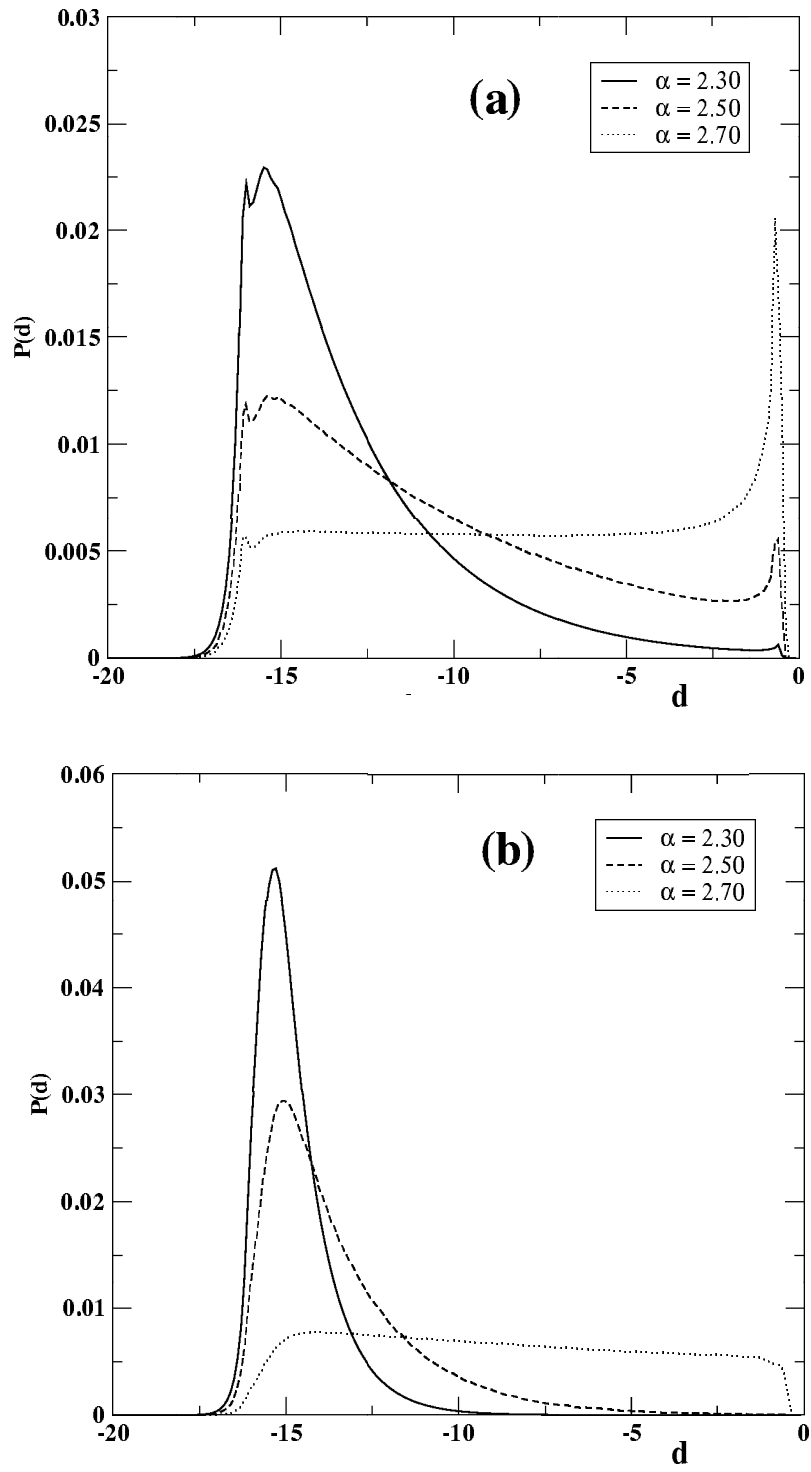

FIG. 9: Probability distribution of log-shadowing distances for the map (a) B and (b) C, for different values of $\alpha$ and with addition of random kicks of strength $\delta_\perp = 10^{-16}$ along the transversal direction.

We define the shadowability time as the interval it takes for the pointwise shadowing distance to grow until $|z_n| > z_A = 0.001$, which is a rather arbitrary choice but with little overall effect on our numerical results. The distribution of shadowing times is thus the same as the power-law distribution of laminar intervals $\mathsf{P}(\tau)$ depicted in Fig. 7. This fact can also be derived by integrating the probability distribution for shadowing log-distances $\mathcal{P}(d)$, in order to obtain the probability for a shadowing distance to be greater than $y_A$ [7].

These probability distributions for the shadowing distances and times can be theoretically justified from the statistical properties of finite-time Lyapunov exponents. A pseudo-trajectory starting off but near the invariant subspace will move either toward or away from $\Omega$ for finite time intervals according to the corresponding time-$n$ transversal Lyapunov exponent: $z_{k+n} \sim z_k \exp(n\lambda_\perp(n))$. It follows that the log-shadowing distances satisfy $d_{k+n} \sim d_k + n\lambda_\perp(n)$. When $\Omega$ exhibits UDV, and especially when it is close to the blowout bifurcation, the time-$n$ exponents $\lambda_\perp(n)$ act as random variables, what makes for an additive random process, with a diffusion rate given by the dispersion of the time-$n$ exponents. Since, in general, the distribution of $\lambda_\perp(n)$ is such that there is a different amount of positive and negative values, we obtain a biased random walk, with a drift towards a reflecting barrier, and which turns to be equal to the average transversal exponent. The external kicks we add are the reflecting barrier for the biased random walk. The distributions for shadowing distances and times may be thus obtained as equilibrium solutions of a diffusion equation taking into account these effects [7, 21].

## VI. CONCLUSIONS

We reported in this paper the parametric evolution of unstable dimension variability (UDV) in strongly non-hyperbolic dynamical systems of low dimensionality and having invariant subspaces possessing a chaotic set. Two key situations are focused in this investigation: the onset of UDV and its maximum intensity at the blowout bifurcation. In the former case, some unstable periodic orbit embedded in the chaotic set loses transversal stability, whereas in the latter, the chaotic set as a whole loses transversal stability. The continuous dependence of UDV on a bifurcation parameter can be described, in periodic-orbit theory, by the relative contribution of the unstable periodic orbits (UPOs) with a different number of unstable directions. These contributions have been numerically tested through the computation of the transverse finite-time Lyapunov exponents and a contrast measure that quantifies the relative contribution of the UPOs.

The sign of the transverse finite-time exponents (which fluctuate about zero when UDV sets in) indicates that a trajectory of finite length is being pushed towards or away the invariant chaotic set on average. According to the periodic orbit theory, the abundance of positive exponents is related to the relative contribution of saddle and repeller UPOs embedded in the chaotic set. At the blowout bifurcation these contributions would match each other, verified for our models by using the computationally largest possible periodics for the UPOs.

Moreover, in our analysis we deal with systems for which there may be UPOs with a larger number of unstable directions. For such systems, we found that the relation (22) between natural measure and the eigenvalues of UPOs is still valid, provided we stick to the original meaning, given in Ref. [3], for the expanding eigenvalues of the UPOs *with unstable directions embedded in the invariant chaotic set, provided the latter is hyperbolic.* In many systems, this is always the case (e.g. in Refs. [4] and [11]), and the difference in the adopted definition of expanding eigenvalue is immaterial. On the other hand, in the systems we deal with in this paper, exhibiting UDV for some parameter values, there may be subtle but important differences due to the possible existence of unstable directions off the chaotic invariant set. In any case, the agreement between periodic orbit theory and numerical experiments can be as good as one wishes, provided we choose UPOs with period large enough.

We paid special attention to the mechanism underlying the onset of UDV which, in this work, differ qualitatively from that that holds for processes of the riddling bifurcation type [17] in two basic aspects. Firstly, for systems undergoing a transition *via* riddling bifurcation, the onset of UDV occurs through the loss of transversal stability of a low-period orbit embedded in the chaotic invariant set. In the mechanism we proposed in this paper, however, as we approach the onset of UDV the period of the orbits losing transversal stability goes to infinity. In the second place, in the riddling bifurcation scenario the onset of UDV involves the collision of three unstable periodic orbits: one transversely stable orbit embedded in the invariant chaotic set, and two other transversely unstable orbits outside the invariant set, such that there are unstable orbits existing *before the onset of UDV*. On the other hand, for the kind of transition shown in this paper, the onset of UDV is followed by the creation of two transversely stable orbits *after the onset of UDV*.

We have also considered the intermittent behavior of trajectories off but very close to the chaotic set. This intermittency is induced by UDV in the sense that the trajectory experiences bursting phases which steer them far away from the chaotic set whenever influenced by transversely unstable UPOs, but can return to the vicinity of the chaotic set under the chief influence of transversely stable UPOs. We obtained that, just after the blowout bifurcation, the probability distribution of inter-burst intervals scales according to the universal 3/2 power-law characteristic of an on-off intermittency in the vicinity of the blowout bifurcation. These findings are according to previous results for UDV in coupled chaotic map lattices presenting transitions to non-synchronized behavior mediated by a blowout bifurcation [18, 31]. In terms of the shadowability of chaotic trajectories, we obtained an exponential distribution for both the log-shadowing distances and a power-law scaling for the corresponding shadowing times, in agreement with other systems studied and also with a theoretical model based on a biased random walk, with a reflecting barrier [7, 21].

To summarize, there are two key points to be emphasized in this paper. Firstly we reinterpreted the two-dimensional map studied in Ref. [11] in terms of UDV. While previous analyses focus on the blowout bifurcation, we also considered the onset of UDV, which occurs through a new mechanism: in the only case described in the literature the onset of UDV occurs through the loss of transversal stability of a fixed point [17], whereas in our case, as we approach the onset of UDV, the period of the orbits losing transversal stability tends to infinity (although this fact seems to depend on the chaotic dynamics in the invariant set). In the second place, we provided another higher-dimensional example which still allows for an analytical treatment for the counting of UPOs embedded in the invariant chaotic set. After describing for both systems the mechanism responsible for the onset of UDV and the point where it is the most intense, we considered the statistical behaviour of FTLE in order to show the observable consequences of the parametric evolution of UDV. The statistical quantities there obtained, the mean and variances, are useful to determine the time during which we expect shadowing trajectories to exist. Moreover, we found that the fraction of positive FTLE coincides, within the numerical accuracy, with the fraction of transversely unstable UPOs, what confirms the usefulness of the periodic orbit theory to describe UDV in a quantitative way. Although we have used low-dimensional model systems to draw these conclusions, essentially the same features are present in higher dimensional dynamical systems. In particular, the systems we used presents the essential features of the synchronization (and transversal) dynamics of spatially extended systems like coupled map lattices or oscillator chains.

## VII. ACKNOWLEDGMENTS

This work was made possible with partial financial support of the following Brazilian government agencies: CNPq, CAPES, and Fundação Araucária (Paraná). The authors would like to acknowledge Dr. A. M. Batista for discussions and useful suggestions.

### APPENDIX: Computing the first orbits to lose transversal stability

We argued, in Section IV, that the onset of UDV occurs when the first UPO embedded in the chaotic invariant set loses transversal stability. In this Appendix we show, using a linear stability analysis, that the first orbit to lose transversal stability in model **B** has a diverging period and does so just after $\alpha = \alpha_c = 1.0$. A similar analysis holds for model **C** and will not be presented here.

Defining the product for all points of a period-$p$ orbit $x_p(j)$

$$X_p(j) = \prod_{i=0}^{p-1} \left( 2^i x_p(j) \bmod 1 \right), \qquad (31)$$

we can write the eigenvalue of this orbit along the transversal direction as

$$\xi_\perp^p(j) = \alpha^p X_p(j), \qquad (32)$$

such that this UPO will be transversely unstable (stable) if $\xi_\perp^p(j) > 1$ ($< 1$). Since $0 \leq x < 1$, it turns out that $X_p(j)$ is always less than unity and there are no transversely unstable UPOs when $\alpha \leq \alpha_c = 1$. We are interested to find the period of the first orbit to lose transversal stability at $\alpha_c = 1$, hence we need to compute the value of $\alpha_{first}(p)$ for which the first period-$p$ loses transversal stability and consider the value of $p$ for which $\alpha_{first}$ approaches $\alpha_c = 1$.

We defined in Section IV, $\alpha_{first}(p)$ as the smallest value of $\alpha$ satisfying $\alpha^p X_p(j) > 1$ for some period-$p$ UPO, or

$$\alpha_{first}(p) > (X'_p)^{-1/p}, \qquad (33)$$

where $X'_p = \sup\{X_p(j)\}$. The orbit which initiates at the point $x_p(j = 2^{p-1} - 1 \equiv J)$ is the only one possessing the remaining $p - 1$ points greater than $1/2$, hence $X'_p = X_p(J)$. There follows that (where each factor is to be understood as $mod\,1$)

$$X'_p = \left(\frac{2^{p-1} - 1}{2^p - 1}\right)\left(\frac{2^p - 2}{2^p - 1}\right)\left(\frac{2^{p+1} - 2^2}{2^p - 1}\right)\left(\frac{2^{p+2} - 2^3}{2^p - 1}\right)\cdots\left(\frac{2^{2p-2} - 2^{p-1}}{2^p - 1}\right) \quad (34)$$

$$= \prod_{i=1}^{p}\left(1 - \frac{2^{p-i}}{2^p - 1}\right) = \prod_{i=1}^{p}\left(1 - \frac{1}{2^i}\frac{2^p}{2^p - 1}\right), \quad (35)$$

in such a way that

$$\alpha_{first}(p) > \left[\prod_{i=1}^{p}\left(1 - \frac{1}{2^i}\frac{2^p}{2^p - 1}\right)\right]^{-\frac{1}{p}}. \quad (36)$$

Let $a_i$ be the $i$th term of the product above,

$$a_i = 1 - \frac{1}{2^i}\frac{2^p}{2^p - 1}.$$

Since $a_i \to 1$ when $i$ (consequently $p$) goes to infinity, Eq. (36) furnishes a convergent result. Since the $a_i$ are always strictly positive and $a_i < a_{i+1} < 1$, this product converges for some $\kappa_p$ such that $0 < \kappa_p < 1$ and Eq. (36) is rephrased as

$$\alpha_{first}(p) > \kappa_p^{-\frac{1}{p}} \qquad (0 < \kappa_p < 1). \quad (37)$$

For large periods, since $a_p \approx 1$, there follows that $\kappa_{p+1} \approx \kappa_p$ and

$$\alpha_{first}(p+1) < \alpha_{first}(p),$$

meaning that orbits with longer periods lose transversal stability earlier than orbits with shorter periods, as $\alpha$ increases. We remark that in other models, like that considered in Ref. [17], the orbits with lowest periods lose transversal stability first. In the $p \to \infty$ limit we find that

$$\lim_{p\to\infty} \alpha_{first}(p) > 1, \quad (38)$$

and the onset of UDV, at $\alpha = 1 + \delta$ with $\delta \to 0$, corresponds to the loss of transversal stability of an infinite period orbit.

---

[1] P. Holmes, Phys. Rep. **183**, 137 (1990).

[2] J.-P. Eckmann and D. Ruelle, Rev. Mod. Phys. **57**, 617 (1985).